# Effect of gallium doping on bubbling and helium retention in aluminum exposed to low-energy helium plasma

Yao Xiao[a,b], Jiliang Wu[a], Shuoxue Jin[b], Chunli Jiang[a], Kai Gu[a], Lei Li[b], Yamin Song[b], Changan Chen[a], Xingzhong Cao[b,*], Xiaoqiu Ye[a,*],

[a] Science and Technology on Surface Physics and Chemistry Laboratory, Mianyang, 621907, China
[b] Institute of High Energy Physics Chinese Academy of Sciences, Beijing, 100049, China

**Abstract:** Surface bubbling and helium (He) retention of pure aluminum(Al) and a solid solution of aluminum-gallium(Ga) alloy (Al-1.8 at.% Ga) exposed to low-energy (50 eV) He plasma has been investigated with the fluence of $1.8 \times 10^{24}$ He/m$^2$ at room temperature (RT). Surface morphology observations show that after irradiation, the average size of He bubbles on Al-1.8 at.% Ga alloy is around 1.8 μm, smaller than that on pure Al (~ 2.5 μm). Ga doping noticeably increases the area density of the bubble. The thermal desorption spectroscopy (TDS) shows that the release amount of He in Al-1.8 at.% Ga alloy is $1.77 \times 10^{21}$ He/m$^2$, nearly three orders of magnitude higher than that in pure Al, whlie the He desorption peak of Al-1.8 at.% Ga alloy is 480 K, much lower than 580 K of pure Al. The results of slow positron annihilation spectroscopy (SPAS) indicate that the vacancy type defects were introduced by He plasma irradiation; and lattice distortion caused by Ga doping plays an important role in determining surface bubbling and He retention characteristics of Al-1.8 at.% Ga.

**Key words**: Al-Ga alloy, SPAS, TDS, high-flux He plasma

## 1. Introduction

Aluminum(Al) is an important material in nuclear reactors due to its low atomic number and low radiation-induced radioactivity[1]. However, Al will become brittle after being introduced a large amount of helium(He) atoms by high-flux density α particles or neutron irradiation, which ultimately reduces the service life of Al [2]. Therefore, in view of safety and economic benefits of the nuclear facilities, it is important to improve the resistance of helium embrittlement in structural materials[3].

*Corresponding author. Science and Technology on Surface Physics and Chemistry Laboratory, Mianyang 621907, China.
E-mail addresses:xiaoqiugood@sina.com (X.Q. Ye), caoxz@ihep.ac.cn (X.Z.Cao).

As we know, He atoms are usually not compatible with any metal atoms, as a rule they sink as $He_mV_n$ or helium bubbles, resulting in the degradation of materials performance such as radiation swelling[4] and surface erosion[5-10]. It has been reported that the alloying elements decrease the binding energy between helium and a vacancy type cluster in tungsten and molybdenum [11]. Zhang Feifei, et al. investigated the behavior of He in $Al/B_4C$ metal matrix composites by transmission electron microscope (TEM) [7]. They found that the He bubbles in Al were much larger than that in $B_4C$. The covalent and ionic bonds in $B_4C$, may be susceptible to ionization induced radiation damage. Since ionization may induce substantial displacements and enhance helium diffusion in $B_4C$, its effect cannot be neglected. The pre-existing defects may act as strong traps for the mobile helium atoms. Ionization and nuclei displacement damages cause the helium atoms to move to pre-existing defects, and promoted the formatin of the helium bubble strings. In addition, the bubbles at the grain boundaries and their vicinity in Al are also faceted. And the diffusion coefficient of the bubble at the grain boundary is larger than that within the grain[12,13]. We investigated the effect of Fe and C doping on the thermal release of helium from Al [3]. The results show that proper fluence of Fe and C doping would lead to the retardation of the release of helium from Al, while excessive fluence of Fe and C doping would result in more desorption peaks and the release of helium in lower temperature ranges. Fe and C doping have different influence on the release of helium from Al, and the difference is related with the secondary phases forming in the samples. Soria[10] found that in a solid solution of Al-wt.4%Cu no He bubbles were observed following irradiation with a 20 keV He ions at room temperature. However, after subsequent annealing at 250 ℃, a homogeneous distribution of He bubbles was observed in the bulk, together with larger bubbles located on the interfaces between the matrix and the equilibrium θ phase ($Al_2Cu$) precipitates.

Both grain boundaries and alloying elements (usually, in the form of precipitates or secondary phases) play an important role on the behavior of He in materials. As we know, Ga can solubilize in Al to form substitutional solid solution. Ga doping in Al can avoid the formation of precipitates or secondary phases. And if pure Al and Al-Ga alloy have the similar grain sizes, it will be more conducive to the understanding of the intrinsic mechanism of the effect of doping elements on the He behavior of Al.

In present work, we chose Al-1.8 at. % Ga alloys with the similar grain sizes to that of pure Al to investigate the effect of Ga doping on the He behavior in Al. Both Al and Al-1.8 at. % Ga were exposed to He plasma with 50 eV/He$^+$ at room temperature to the fluence of $1.8 \times 10^{24}$ He/m$^2$. The effects of Ga doping on the phase composition, morphology, defect distribution and the He release behavior of Al were investigated.

## 2. Experimental

The samples used are high purity (99.99%) Al and Al-1.8 at.% Ga alloy. Al-1.8 at.% Ga alloy were prepared by cooling the alloys to room temperature after melting Al with Ga (99.99% purity) at a high temperature(> 950 K). To prevent the segregation of Ga, the samples were annealed at 793 K for 80 hours.

To get the similar grain size, Al and Al-1.8 at.% Ga alloy experienced different machining and annealing processes. The original Al and Al-1.8 at.% Ga alloy were compressed by 3% and 1%, and then annealed at 648 and 673 K for 30 minutes, respectively. And the grain size was observed by using an optical microscope. The annealing process not only contributes to grain growth, but also reduces stress and defects. The corrosion solution (HCl:HNO$_3$:HF:H$_2$O =9:3:3:5 in mole ratio) was used to erode the dislocations in samples.

Fig.1 shows the average grain size of both samples is about 300 μm. There are also some particular corrosion pits, which are the signs of dislocation outcrops [14,15]. Because of the low contrast in the figure, the resolution of the dislocation outcrop is poor, and the details can be seen in SI. It can be found that the dislocation outcrops in the Al-1.8 at. % Ga alloy were more than that in pure Al. This is also confirmed by

XRD as shown in the supplementary materials (SI).

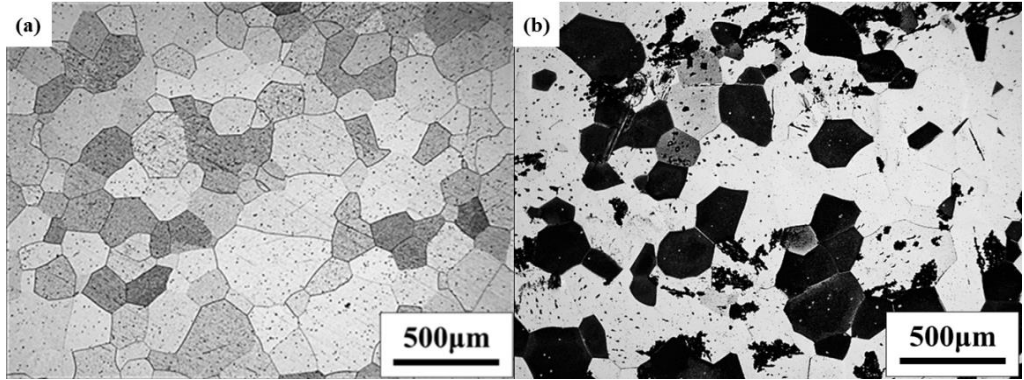

Fig.1 The metallographic diagram of Al (a) and Al-1.8 at.% Ga alloy (b)

He plasma exposure was performed for Al and Al-1.8 at.% Ga alloy in the linear experimental plasma system at Science and Technology on Surface Physics and Chemistry Laboratory. remove the surface contaminants and oxide layers, the samples were sputtered with 146mA He plasma for 10 min before He irradiation at RT. The He ion energy and the beam flux were 50 eV/He$^+$ and $3.7\times10^{20}$ ions/(m$^2$ s), respectively. The ion fluence was $1.8\times10^{24}$ He/m$^2$. During irradiation, the temperature of the targets were kept at room temperature.

The slow positron annihilation spectroscopy (SPAS) measurement was performed at room temperature by using energy-variable slow positron beam equipment in the Institute of High Energy Physics Chinese Academy of Sciences [16]. Detection of defects in irradiated and unirradiated Al and Al-1.8 at.% Ga alloy by coincidence doppler broadening(CDB) of SPAS. Positrons were generated by $^{22}$Na radiation source and then moderated by W foil to obtain slow positrons with energy ranging from 0.18 keV to 20.5 keV. The injection depth of the positron can be estimated with the Eq.1, in which $x$ is expressed in nanometers

$$x = \frac{40}{\rho} \times E^{1.6} \qquad (1)$$

where E is the kinetic energy of the incident positron (keV) of the slow positron, ρ the density of material in units of g/cm$^3$($\rho_{Al}$=2.699 g/cm$^3$). The $S$ parameters are defined as the ratios of the counts in central low momentum area (510.2-511.8 keV) in doppler broadening (DB) spectra which is sensitive to defects. When the number density of defects increases, the *S* parameter increases. The *W* parameters are defined as the ratios of the counts in two flanks high momentum regions (514.83-518.66 keV and 503.34-507.17 keV) in the DB spectra to the total counts [17].

Surface morphology of the samples was observed by using a scanning electron microscope (SEM). Thermal desorption spectroscopy (TDS) was used to get data on He trapping and retention. During the TDS scan, the sample was held in a quartz tube with a pressure below $10^{-5}$ Pa, and was heated up to 953 K with a ramping rate of 10 K/min. During the heating, a quadruple mass spectrometer (QMS) was utilized to monitor the indicators of mass 4 (He). QMS signals for mass 4 (He) are calibrated using a helium leak with a known leak rate before every TDS measurement. The relative He sensitivity of QMS was determined previously and was assumed to be stable (details are given in [18]).

## 3. Results and discussion

### 3.1. Surface bubbling

Fig.2 a and c shows that the initial surface morphology of Al and Al-1.8 at. % Ga alloy are smooth with only a small amount of scratches. After irradiation, many He bubbles appeared on the surface of Al and Al-1.8 at. % Ga, as shown in Fig.2c and Fig.2d. However, the size of He bubbles are not uniform. In some zones, small-sized bubbles are densely packed, while in other zones, large-sized bubbles are formed with a relatively dispersed distribution. In addition, it seems that Ga doping significantly increases the area density of the bubble. And the average size of He bubbles on Al-1.8 at.% Ga alloy is around 1.8 μm, smaller than that on pure Al(~ 2.5 μm)( After 10 keV He implantation, the bubble on surface of the material can still reach the size of μm, which can be seen in SI). Some large bubbles were broken, causing surface flaking or leaving the holes on the surface of Al and Al-1.8 at. % Ga alloy.

This kind of phenomenon was seldom reported before. We think that it may be due to the non-diffusing nucleation of He [6]. Different from He ion plantation, He plasma with the low energy and high flux can not only sputter clean the surface, but also accumulate to a high concentration of He in a very short time. When the concentration of He reaches a critical concentration that can be randomly combined, the surface is bubbling [6]. As the dose is further increased, the surface begins to repeat peeling [9,12], which is shown in Fig.2b. The non-uniform distribution of the bubbles in Fig.2 may be due to the aggregation of He at different defects such as dislocations, grain boundaries and so on, since these defects are strong capturing traps of He [19].

Note specially, Ga doping reduced the average size of He bubbles and increased the density of the bubble. It may be attributed to the more dislocations in Al-1.8 at. % Ga, since dislocations are usually the nucleation centers of the bubble, and thus reduced the segregation of the bubbles. We will discuss this further later.

In addition, it was found that the bubble cracking in Al-1.8 at.% Ga alloy was more serious than in Al ( see SI). This is because the mechanical properties of Al-1.8 at.% Ga alloy are inferior to those of Al [20,21].

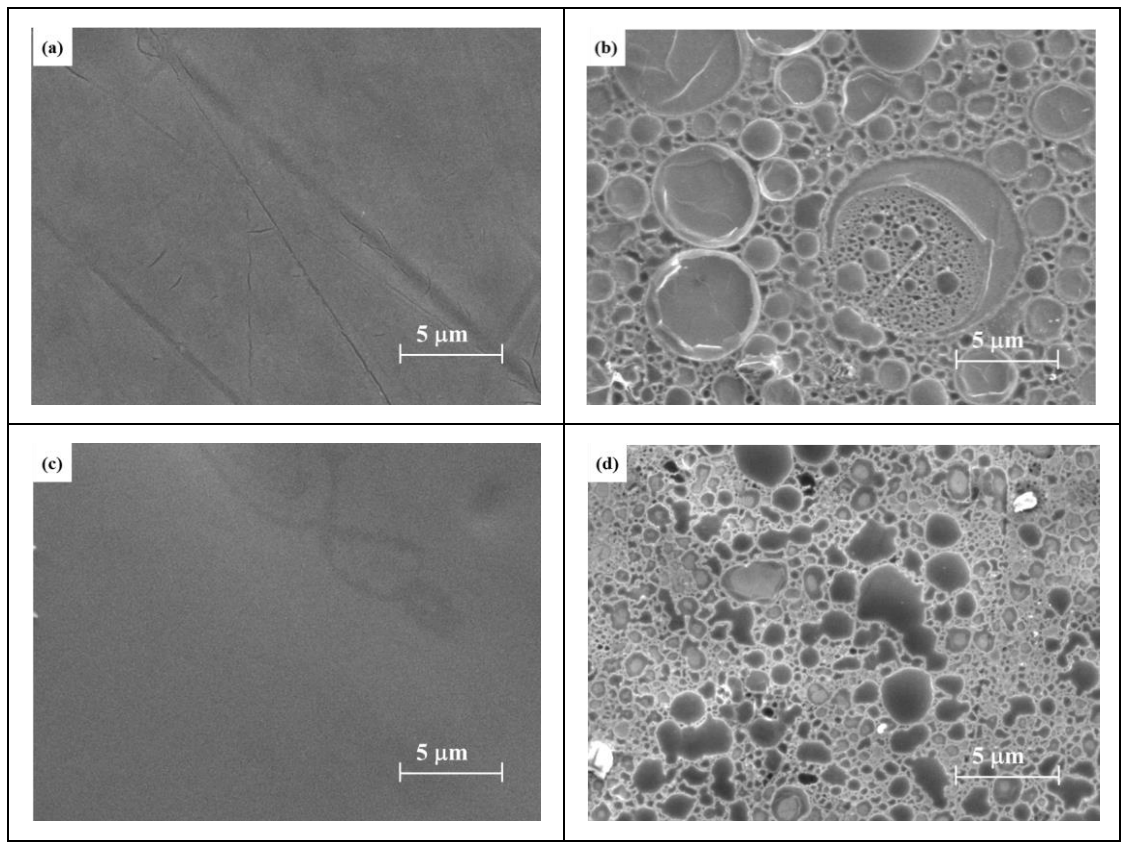

Fig.2 SEM of Al and Al-1.8 at.% Ga alloy before and after He irradiation
(a) Al before He irradiation (b) Al after He irradiation (c) Al-Ga before He irradiation (d) Al-Ga after He irradiation

### 3.2. Helium desorption

Fig.3 shows the TDS spectra of He in Al and Al-1.8 at.% Ga alloy. It can be seen that the He desorption peak of Al-1.8 at.% Ga alloy is 480 K, much lower than 580 K of pure Al. And note specially, the release amount of He in Al-1.8 at.% Ga alloy is $1.77 \times 10^{21}$ He/m$^2$, nearly three orders of magnitude higher than that in pure Al. It seems that Ga doping greatly has an impact on the thermal release behavior of He in Al.

The difference between Al and Al-1.8 at. % Ga alloy in TDS may be attributed to the Al lattice distortion caused by Ga doping, since more dislocations can be found in Al-1.8 at. % Ga, as shown in Fig.1 and confirmed by XRD in Fig.3 in SI. Dislocations play a dominant role in He bubble nucleation[22], and also promote helium desorption from He$_n$V$_m$ clusters[23].

It can be also seen that there are several satellite peaks around the main peaks of Al and Al-1.8 at. % Ga. It indicates that there are different defects that trapped He atoms[24]. The release of He in the low temperature region(300-490K) is generally considered to be He released from defects such as grain boundaries, dislocations, and the like which are lower in binding energy with He in Al [9], while the release of He in the high temperature region(490-750 K) can be attributed to the decomposition of He$_m$V$_n$ in Al[3]. Thus, it is necessary to evaluate the types of defects in Al and Al-1.8 at. % Ga. We will discuss this in the following section.

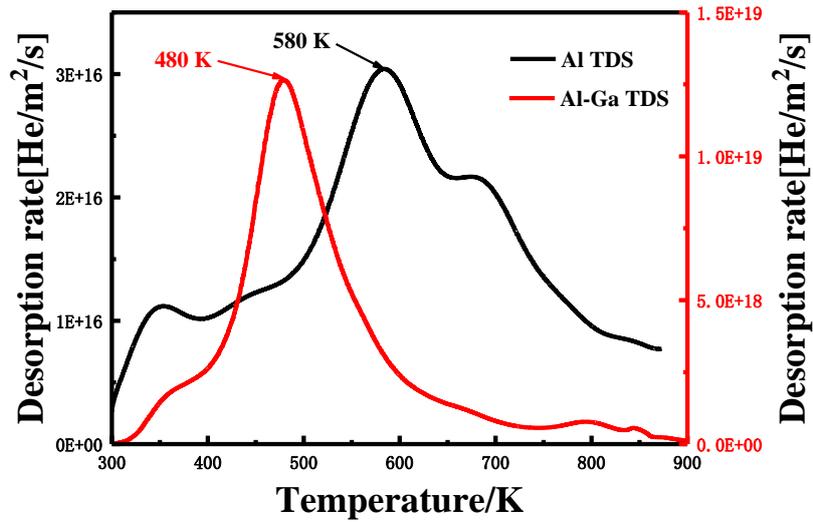

Fig.3 S TDS of Al and Al-1.8 at.% Ga alloy after He irradiation

### 3.3. Vacancy-type defects produced by He plasma irradiation

SPAS is a very effective nuclear spectroscopy method for obtaining the surface structural defects of solid materials and their characteristics [25]. S parameter is sensitive to open volume defects, and W parameter is sensitive to the chemicals in the environment of the annihilation site [17]. In other words, a decrease in the S parameter indicates a reduction in vacancy-type defects density. Furthermore, the combination of S and W parameters can be used to distinguish different kinds of vacancy-type defects. For a certain type of defects, the (S, W) plots appear along a fairly straight line. A large deviation between two (S, W) plots indicates that the positrons are annihilated to produce different kinds of vacancy-type defects [26].

Fig.4 shows the *S* parameter as a function of positron incident depth (*S-D* curve) dependent on defects in Al and Al-1.8 at. % Ga alloy.

For unirradiated samples, the surface and the bulk *S* parameters are not the same. This indicates that the surface condition is not consistent with the internal condition of these samples even after annealing. The S parameter was clustered on the surface, because defects are introduced into the surface due to the mechanical polishing. This kind of detects may be vacancy clusters [27].

For the unirradiated samples, the bulk S parameter of Al-1.8 at. % Ga alloy is much larger than that of Al. It further confirms that there are more vacancy-like defects caused by lattice distortion in Al-1.8 at. % Ga alloy. As we mentioned before, this kind of vacancy-type defects are dislocations. The reasons why these dislocations are introduced by Ga doping rather than by the sample preparation process are detailed in SI. The cause of lattice distortion is that Ga, which mismatched with the matrix lattice, is doped into the Al.

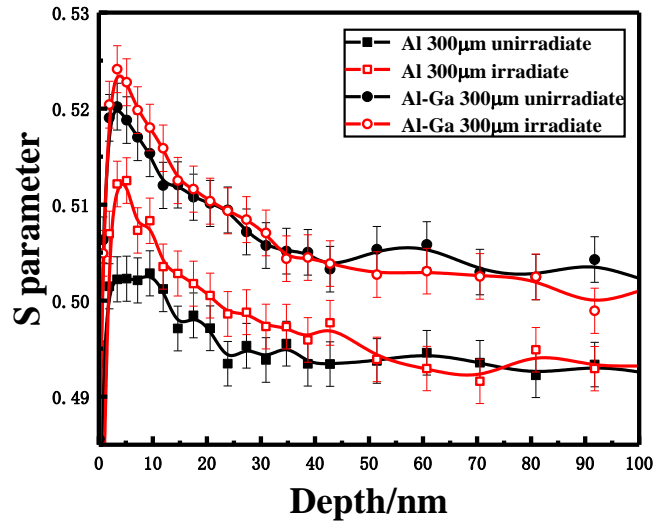

Fig.4 *S-D* curves for Al and Al-1.8 at. % Ga alloy before and after irradiation

The sequence of the creation and annihilation of defects during he- lium ion irradiation process is as follows: defects appear in the helium ion bombardment material; helium ion bombardment material produces temperature effects, which cause defects to annihilation; helium ion and vacancy are combined to form helium-vacancy complex and helium bubbles to reduce the number of vacancies [28].The value of *S* parameter reached the maximum at the depth of ~3 nm, which is similar to the damage depth of ~3 nm from SRIM simulation [5]. In the interior of the material, the S parameters remarkably decreased. As expected, after irradiation, He atoms ($He_m$) would occupy vacancies ($V_n$) and form the $He_mV_n$ complexes at room temperature, which can affect the annihilation of positrons with the electrons in defects [29,30]. However, the helium ion bombardment caused the surface to produce more vacancies. This was confirmed by the larger *S* values of irradiated samples relative to that of the unirradiated, as shown in Fig.4. Moreover, the *S* parameter of irradiated Al-1.8 at. % Ga alloy is still larger than that of irradiated Al. This indicates that defects of the Al-1.8 at. % Ga alloy are more than those in Al.

The slope of the *S-W* line of irradiated Al-1.8 at. % Ga alloy is the smallest, while that of unirradiated Al is the largest. The slope of *S-W* for irradiated Al is larger than that of unirradiated Al-1.8 at. % Ga alloy. These phenomena are consistent with the CDB results. The decrease of the gradient values compared to the not irradiated Al-1.8 at. % Ga alloy was due the lattice distortion caused by Ga doping. The slope of the *S-W* line of unirradiated samples are larger than those of the irradiated samples, because the vacancy-type defects were introduced into the material after He irradiation. The slope of the *S-W* plot could represent mechanism of positron annihilation after trapped, and the *S-W* plot could be used to identify the number of defect types which are trapped by positrons in materials[17]. These say that only one type defect was detected in irradiated samples and unirradiated Al-1.8 at. % Ga alloy by the positrons.

From the above analysis, it can be concluded that the lattice distortion caused by Ga doping plays a dominant role in affecting the behavior of He in Al. The vacancy-type defects caused by the lattice distortion become the nucleation center of $He_mV_n$, which reduces the average size of the bubble and increases the area density of the bubble. Increased defects make it easier for materials to capture He. The generation of new dislocations causes the He to be released into the low temperature zone.

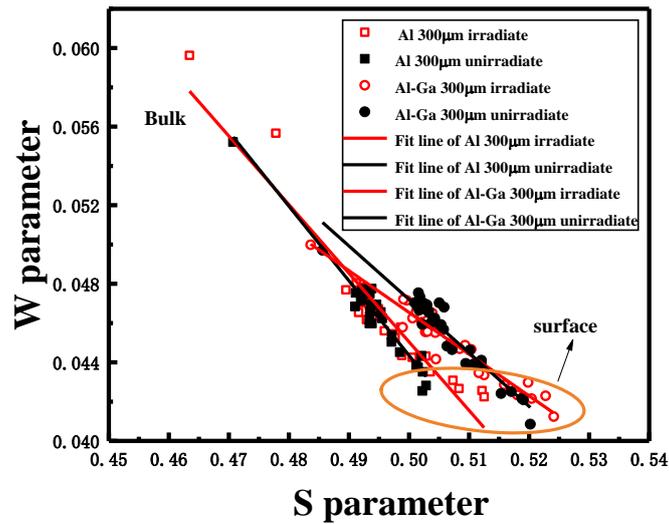

Fig.5 *S* versus *W* plots for Al and Al-1.8 at. % Ga alloy before and after He irradiated

## 4. Conclusion

Al and Al-1.8 at.% Ga alloy were exposed to He plasma with 50 eV/He+ at room temperature to the fluence of $1.8 \times 10^{24}$ He/m$^2$. The He retention and its irradiation damage behavior were investigated. The results show that doping Ga has an obvious impact on the behavior of He in Al:

(1) After the He plasma exposure, the Ga doping significantly increases the bubble density and decreases average size of bubble. However, the surface bubble bursting is exacerbated by mechanical properties of the material weakening.

(2) The release amount of He in Al-1.8 at.% Ga alloy is nearly three orders of magnitude higher than that in pure Al; while the He desorption peak of Al-1.8 at.% Ga alloy is 480 K, much lower than 580 K of pure Al. The He retention and diffusion characteristics of Al-1.8 at.% Ga alloy may be mainly attributed to dislocations and microcracks caused by lattice distortion.

(3) The results of SPAS and TDS indicate that a large number of vacancy-type defects which is induced by doping Ga. The increasing defects/He-trap sites may result in two outcomes, the He bubble nucleation site is increased and He inward diffusion is enhanced, therefore, lead to the mitigation of He-induced bubbling and increase the He retention.

## Acknowledgements

We are grateful to Yitao Yang, Pro. Chonghong Zhang, Yongli Liu and Qingyuan Liu for their discussions. We acknowledge support by National Magnetic Confinement Fusion Energy Research Project (No. 2015GB109002) from Ministry of Science and Technology of China. The authors also appreciate the support from the National Nature Science Foundation of China (No. 21401173).